\documentclass[aps,preprint,groupedaddress]{revtex4-1}

\usepackage{xspace}

\usepackage{notoccite}

\usepackage{url}
\usepackage{hyperref}
\usepackage{ dsfont }

\usepackage{graphicx}

\begin{document}



\title{Particle Camera MX-10 in Physics Education}




\author{Ji\v{r}\'{\i} Kvita}
\email[]{jiri.kvita@upol.cz}
\affiliation{Regional Centre of Advanced Technologies and Materials, Joint Laboratory of Optics of Palack\'{y} University and Institute of Physics AS CR, Faculty of Science, Palack\'{y} University, 17. listopadu 12, 771 46 Olomouc, Czech Republic}

\author{Berenika \v{C}erm\'{a}kov\'{a}}
\author{Nat\'{a}lie Matulov\'{a}}
\affiliation{Grammar School of Nicolaus Copernicus in B\'{\i}lovec, 17. listopadu 526, 743 11 B\'{\i}lovec, Czech Republic}

\author{Jan Po\v{s}tulka}
\author{Daniel Stan\'{\i}k}
\affiliation{Grammar School in Uni\v{c}ov, Gymnazijn\'{\i} 257, 783 91  Uni\v{c}ov, Czech Republic}
\thanks{\it To the memory of Karel Tesa\v{r}, high school teacher from the Grammar School in Uni\v{c}ov.}


\date{\today}

\begin{abstract}
We present several applications of the particle camera MX-10 in different radiation environments, leading to strikingly different observed patterns and consequently particle composition as well as recorded dose. We describe the measurements of the radiation background, cosmic muons direction analysis and the detection of alpha particles in a natural cave. We focus on the analysis of tracks of highly ionizing particles aboard an airplane and compute the dose rate with altitude. We also comment briefly on a test in beams of particles at an accelerator. We discuss the problem of presenting sensitive results on radiation background levels in various environments to general public. Most of the experiments presented here were performed and analyzed by high-school students, confirming the application of the device not only for class-room  demonstrations but also for science projects in physics and IT.
\end{abstract}

\pacs{XYZ}

\maketitle





\section{Introduction}

\begin{figure}[!t]
  \centerline{
    \includegraphics[width=0.800\textwidth]{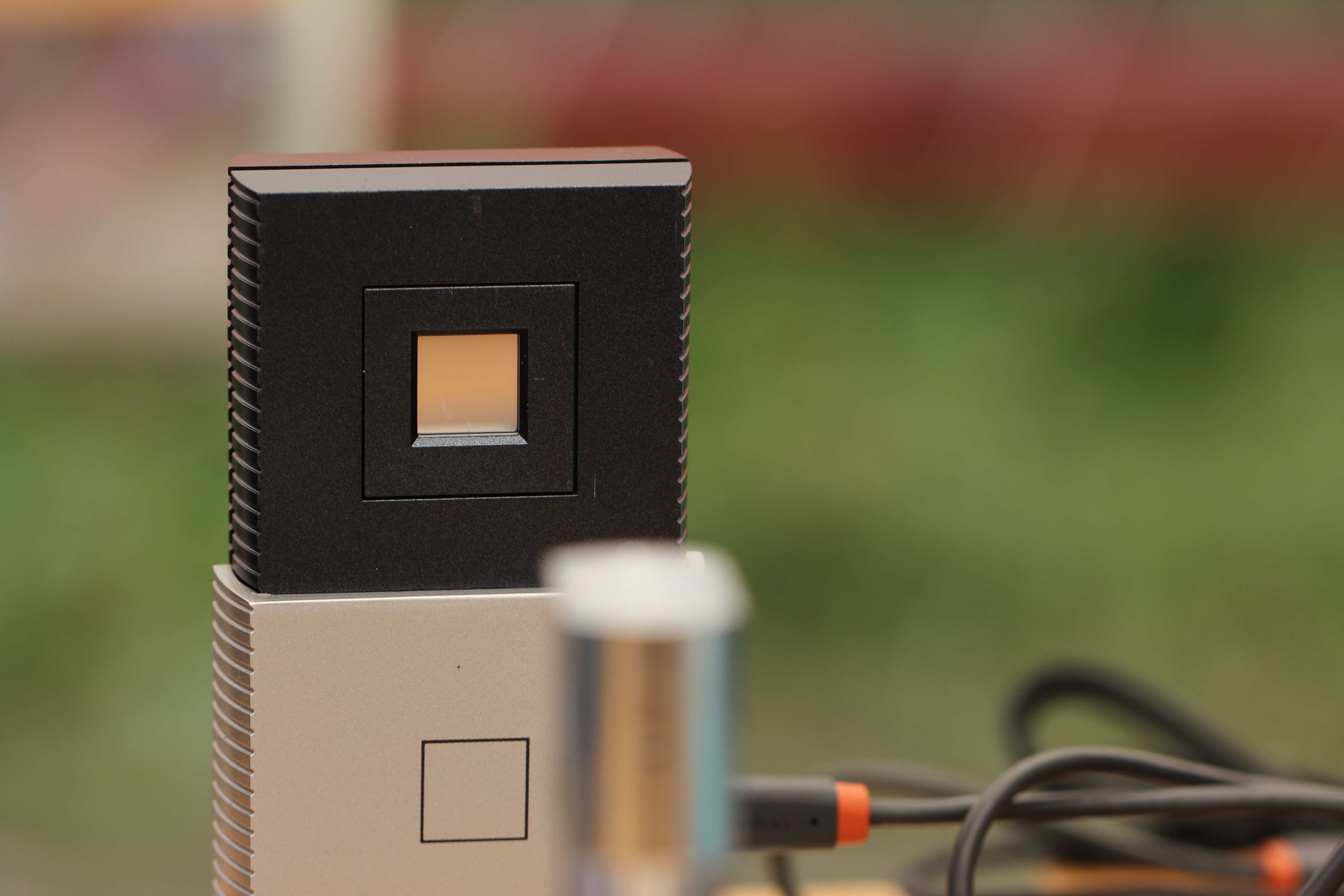}  }
  \caption[]{The particle camera MX-10~\cite{HOLY2006254,VYKYDAL2006112} based on the Medipix2 technology, with an the open cover showing the aluminum-coated sensor.}
\label{fig:camera} 
\end{figure}

The particle camera MX-10 (see Fig.~\ref{fig:camera}) of the Medipix2 family~\cite{HOLY2006254,VYKYDAL2006112} is a professional solid-state particle detector with the capability of energy and time measurements as well as particle identification based on the observed pattern of signaling pixels.

The device is an excellent educational equipment for demonstrating various properties of the ionizing radiation, able to show distinct patters for gamma (few-pixel dots), beta (visibly long and curved tracks), alpha (wider blobs due to the charge sharing between pixels), muons (long straight tracks) or more exotic wide tracks of heavily ionizing particles. We will describe the examples of all these in more detail in corresponding sections together with ideas on how to use the observations to explain key physics features of the conditions to obtain the patterns as well as the processes leading to observing the particles in the first place.

One of the device key features is the possibility to demonstrate the shielding effects of air or a sheet of paper to stop alpha particles, or various metals of different thickness to stop beta and gamma rays. The educational set provided by Czech company JABLOTRON ALARMS includes the shielding material as well as optionally also radiation sources like glass doped with uranium dioxide, metallic  electrode doped by a thorium dioxide, or americium sources modified to primary gamma or alpha sources.

\section{Particle camera in the laboratory}

\subsection{Radiation background}

Laboratory measurements offer many opportunities to analyze the shielding power of various materials like metals of different proton number. School and laboratory $\alpha/\beta/\gamma$ radioactive sources can be used to demonstrate the shielding power in terms of the number as well as energy spectra of detected particles. One can also study several other accessible sources like potassium-rich fertilizers or uranium-oxide doped glass used for art and decoration purposes.
Other sources include dust collected on a paper or cloth filter after vacuum-cleaning a room: it includes solid radioisotopes from the radon decay chain.

\begin{figure}[!h]
  \includegraphics[width=0.450\textwidth,height=0.450\textwidth]{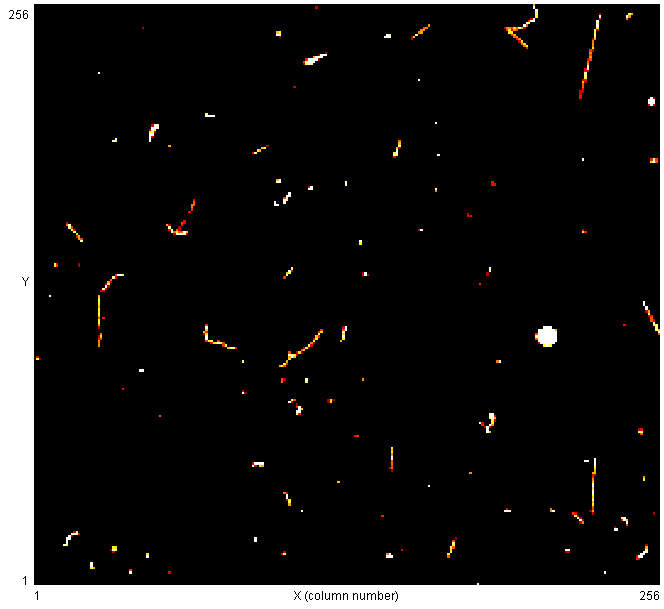}
  \includegraphics[width=0.450\textwidth,height=0.450\textwidth]{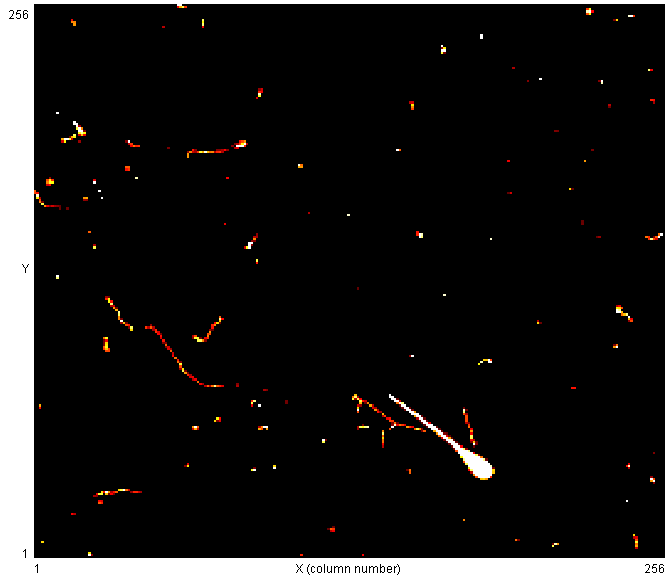} \\
  \caption{Example of a 10 min background exposure with a recorded alpha particle from natural background radiation (left) and a rare track of a highly-ionizing particle at the ground level (right).}
\label{fig:bg_HI}
\end{figure}

But the simplest one is the background itself, dominated by gamma and beta particles, occasionally spiced by alpha particles from the aforementioned and ubiquitous radon, but also more exotic particle species like cosmic muons or even heavy ionizing energetic particles, see~Fig.~\ref{fig:bg_HI}.

\subsection{Cosmic muons direction analysis}

Educationally, a single image of a cosmic muon is a simultaneous proof of time dilatation by Einstein, proof of the existence of the second family of elementary particles, and a proof of radiation of extra-terrestrial origin.

\begin{figure}[!h]
  \includegraphics[width=0.450\textwidth]{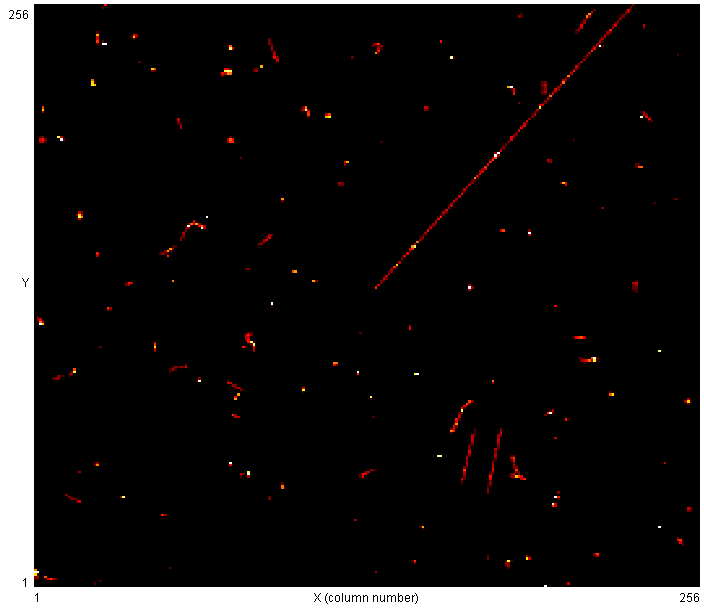}
  \includegraphics[width=0.450\textwidth]{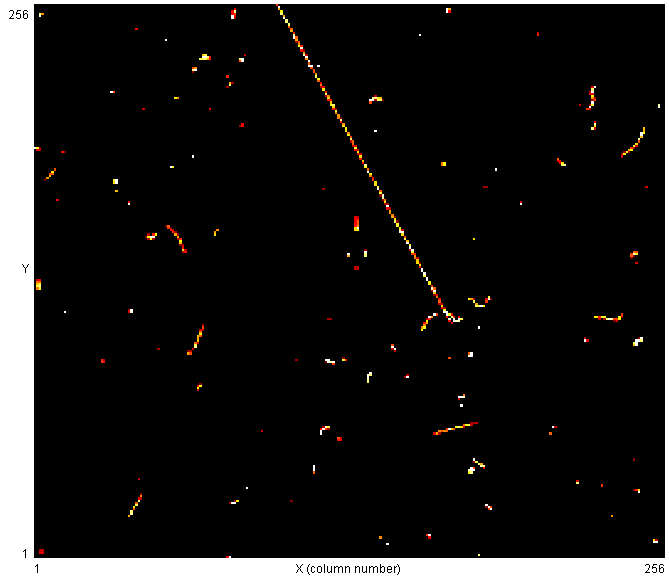}
  \caption{Example of clean tracks of passing cosmic muons recorded in the laboratory during 10 min exposures, also with non-negligible gamma and beta background.}
\label{fig:lab}
\end{figure}

For the analysis of the direction of incoming muons, data have been recorded in frames of 10~min each. During the total exposure of 662 hours, sometimes an accidental highly ionizing particle, perhaps from an extensive cosmic ray shower, was recorded, as seen in Fig.~\ref{fig:bg_HI} (right), with a rate of approximately one per day. In total, 225 lines were found of length of at least 60 pixels, i.e. about one such a candidate per 20~min. Typical events are shown in Fig.~\ref{fig:lab}.

\begin{figure}[!h]
  \includegraphics[width=0.950\textwidth]{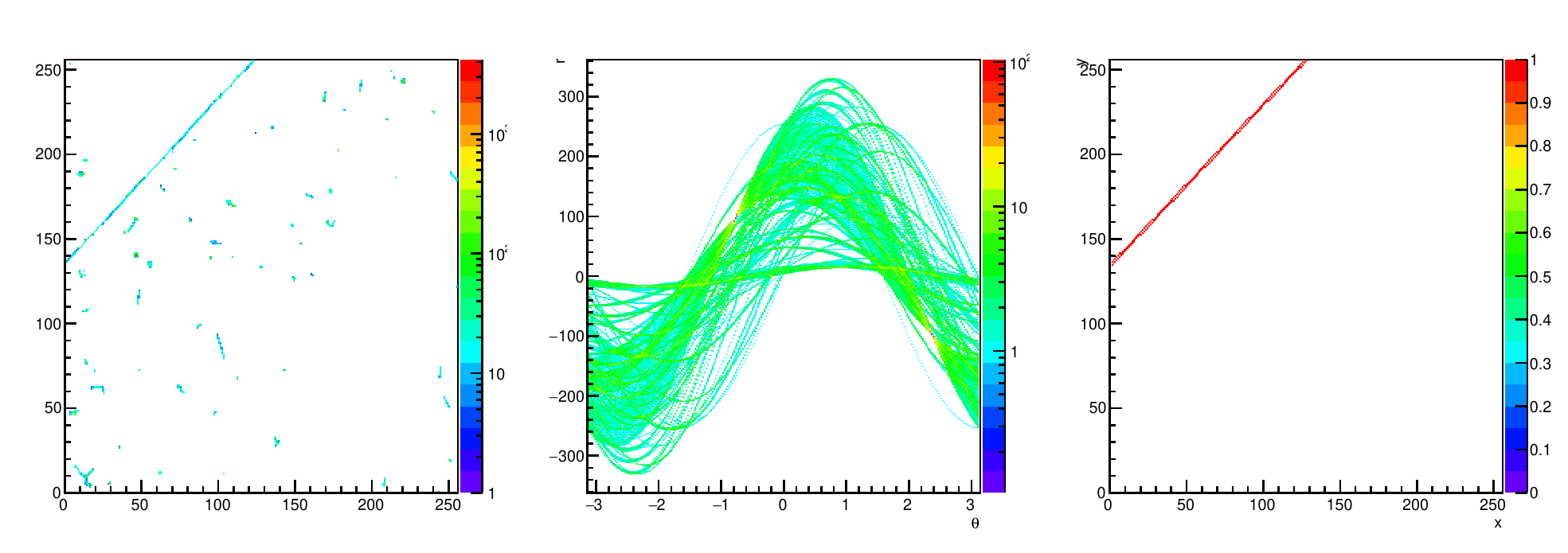}
 \caption{An example of an event with a straight track from a passing cosmic ray muon (left), the Hough-transformed image (middle) in the $\theta-r$ space, and the reconstructed lines (right).}
\label{fig:hough}
\end{figure}

Hough transformation, originally proposed for the analysis of bubble chamber pictures in 1959~\cite{Hough:1959qva}, was used to search for line patters, candidates for the straight muon tracks. A simple, straightforward yet inefficient implementation of the Hough transformation was written in Python to analyze the recorded frames. Only the longest track from each frame was accepted to the analysis. As illustrated in Fig.~\ref{fig:hough}, the algorithm transforms the 2D image from the $x-y$ plane to the $\theta-r$ space, where $r$ is the closest approach of a line at angle $\theta$ w.r.t. the origin. In essence, all lines are tried, pixels along the line are analyzed and the number of pixels with non-zero energy deposit is counted. This number is set as the value of the Hough-transformed histogram in the $\theta-r$ space at coordinates corresponding to the line parameters. Searching numerically for the maxima leads to finding longest lines of given parameters.

\begin{figure}[!h]
  \centerline{
    \includegraphics[width=0.750\textwidth]{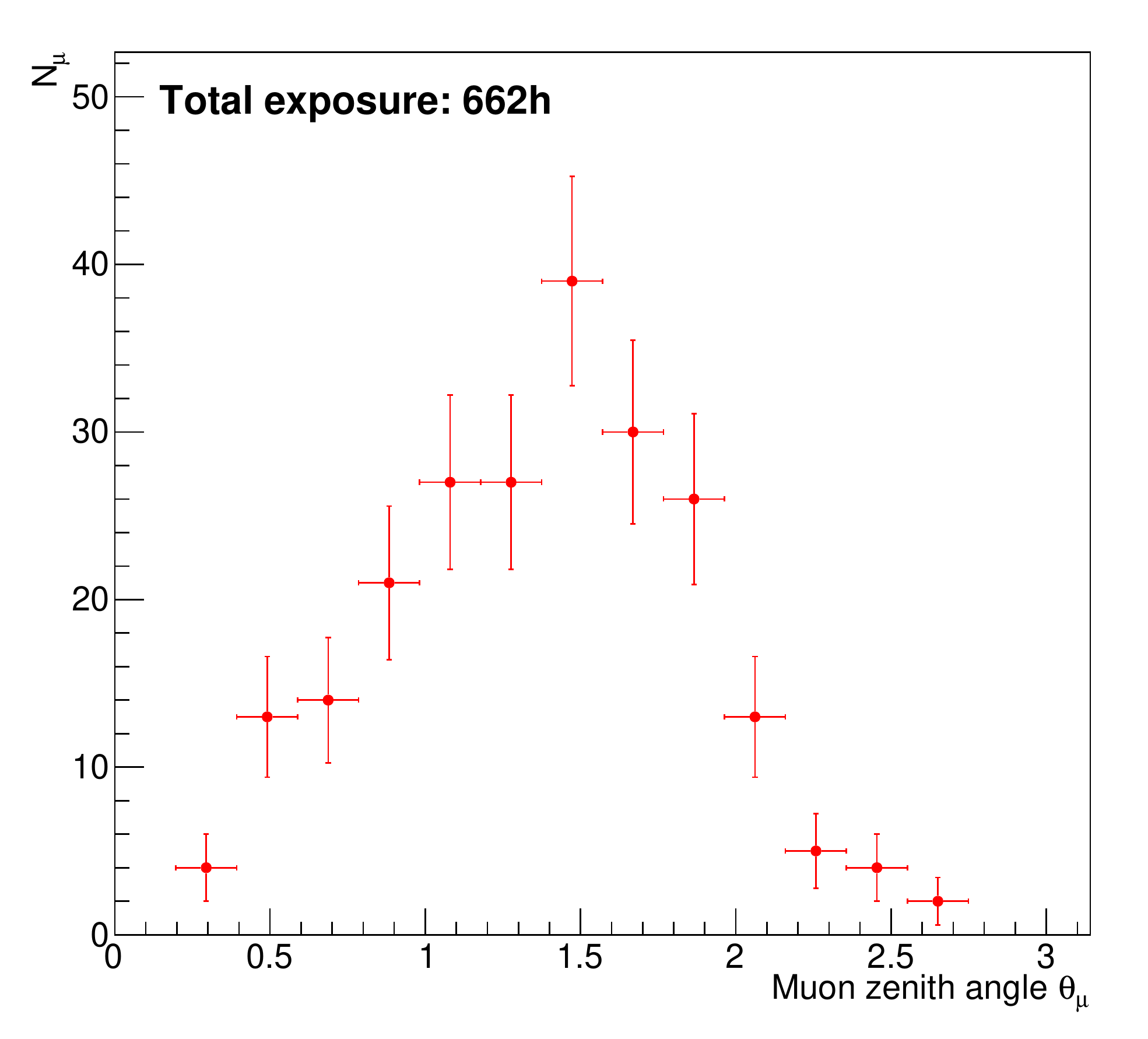}  }
  \caption{Distribution of the zenith angle of the incoming cosmic-ray muons, showing their dominant direction of coming from above.}
\label{fig:zenith}
\end{figure}

The distribution of the azimuthal angle $\theta_\mu$ of the muon candidates in Fig.~\ref{fig:zenith} is peaked around the direction of vertically-incoming muons. It is slightly shifted towards smaller angles, which could be attributed to shielding by the building (the experimental setup was kept close to window, i.e. close to a more open side of a building).

\section{Particle camera in caves}

\begin{figure}[!h]
\centerline{ 
  \includegraphics[width=0.450\textwidth]{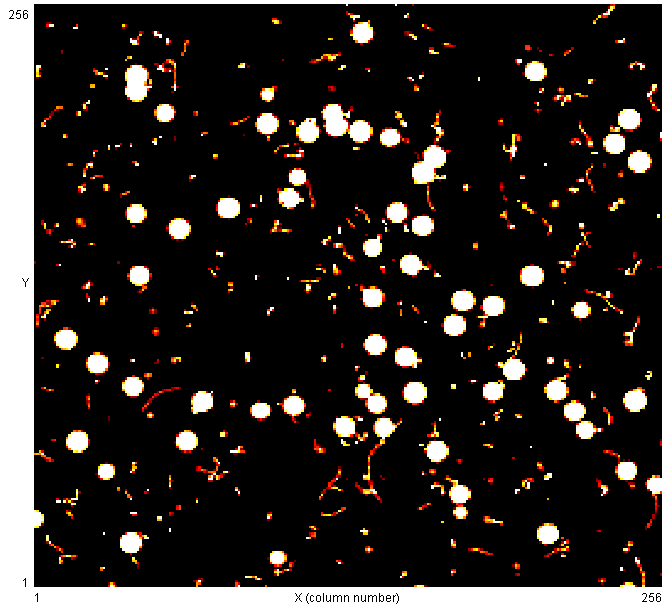}
  \includegraphics[width=0.450\textwidth]{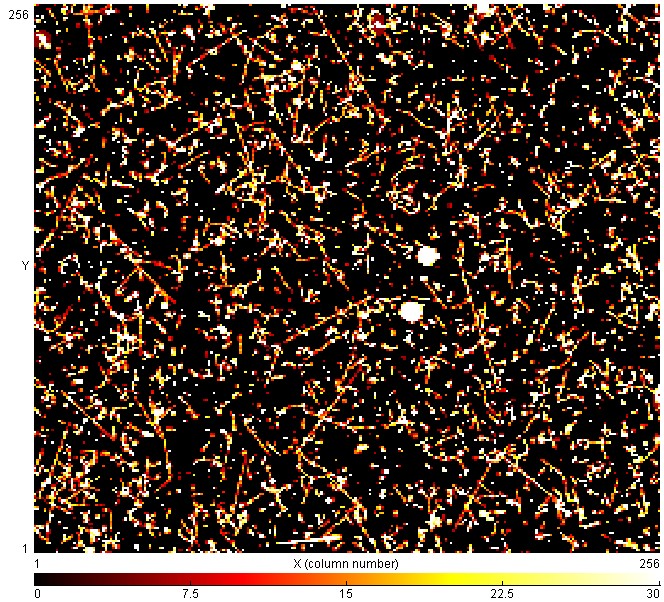}
}
  \caption{Example of a frame of 10~min in a natural cave (left) and of 6~h of natural background (right).}
\label{label:caves}
\end{figure}

\begin{figure}[!p]
\centerline{ 
  \includegraphics[width=0.950\textwidth]{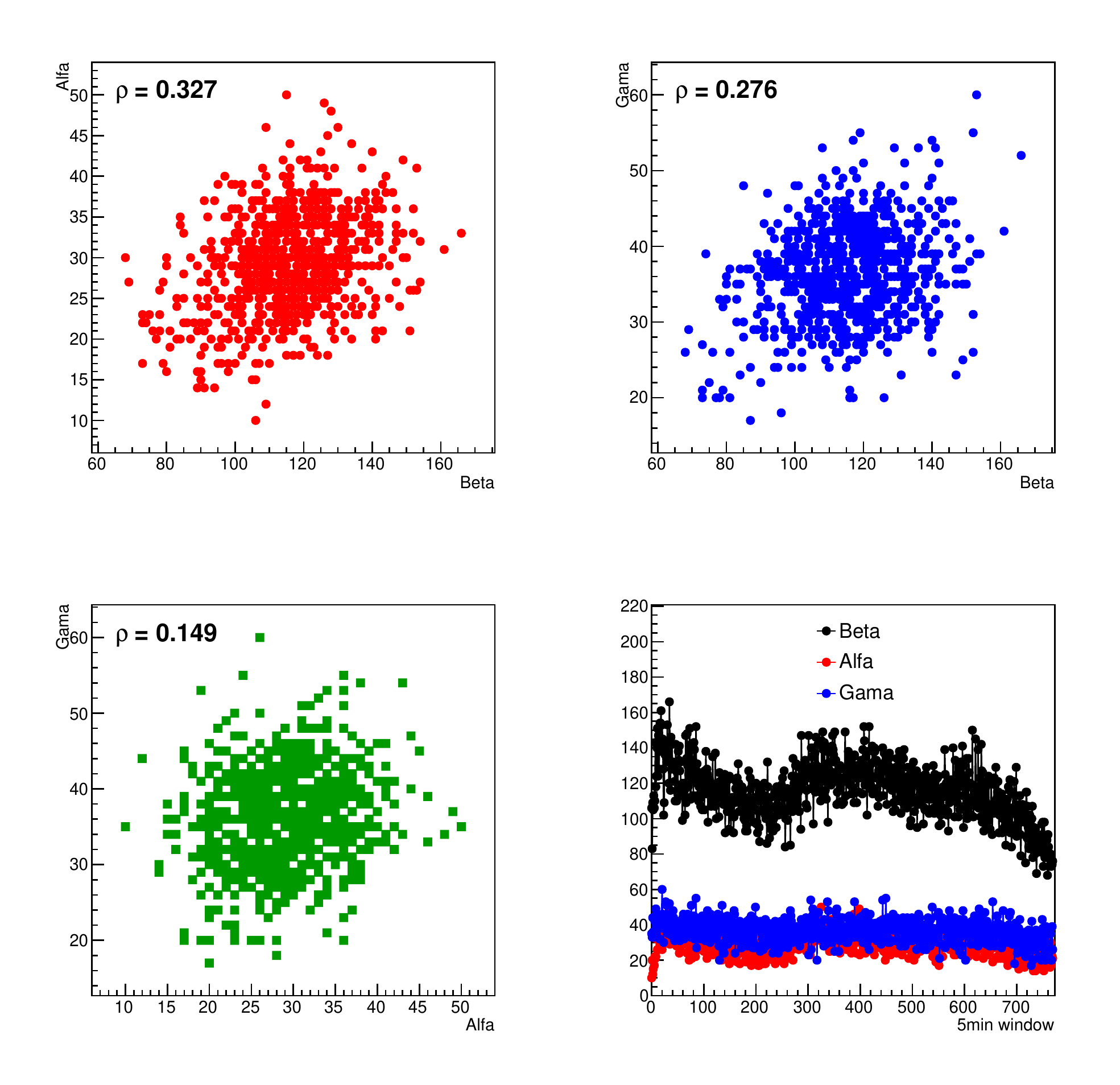}
}
  \caption{Correlations over frames between the observed numbers of particles of different kinds, and their multiplicity over time in 5~min windows measured in a natural cave. Recorded using the MX-10 particle camera, with multiplicities analyzed by a private code.}
\label{fig:corr}
\end{figure}

The camera has been installed for a total of 2 days and 16 hours in a natural cave with an influx of carbon dioxide along a geological crack in limestone bedrock, expected to bring more radon from the depths of the Earth. A typical observed frame is shown in Fig.~\ref{label:caves}, compared to the composition of particles recorded in a room on surface.
Interestingly, in contrast to laboratory measurements, no muons were recorded due to the shielding of the rock and soil, and the frames are visually dominated by alpha particles, although the multiplicities of beta and gamma particles are higher (see Fig.~\ref{fig:corr}).

Further, multiplicities of particles of different kinds were analyzed by a private code written by students, comparing very well to the original SW shipped with the camera, enabling to resurrect data on individual particle multiplicities lost in one set of measurements, where only the total multiplicity was saved to a data file. Correlations over frames between the observed numbers of particles of different kinds are shown in Fig.~\ref{fig:corr}. Students have thus trained themselves in programming, pattern recognition, algorithm development and finally also in statistics, error treatment, and even covariance and correlation estimation.

\begin{figure}[!p]
  \centerline{
    \includegraphics[width=0.750\textwidth]{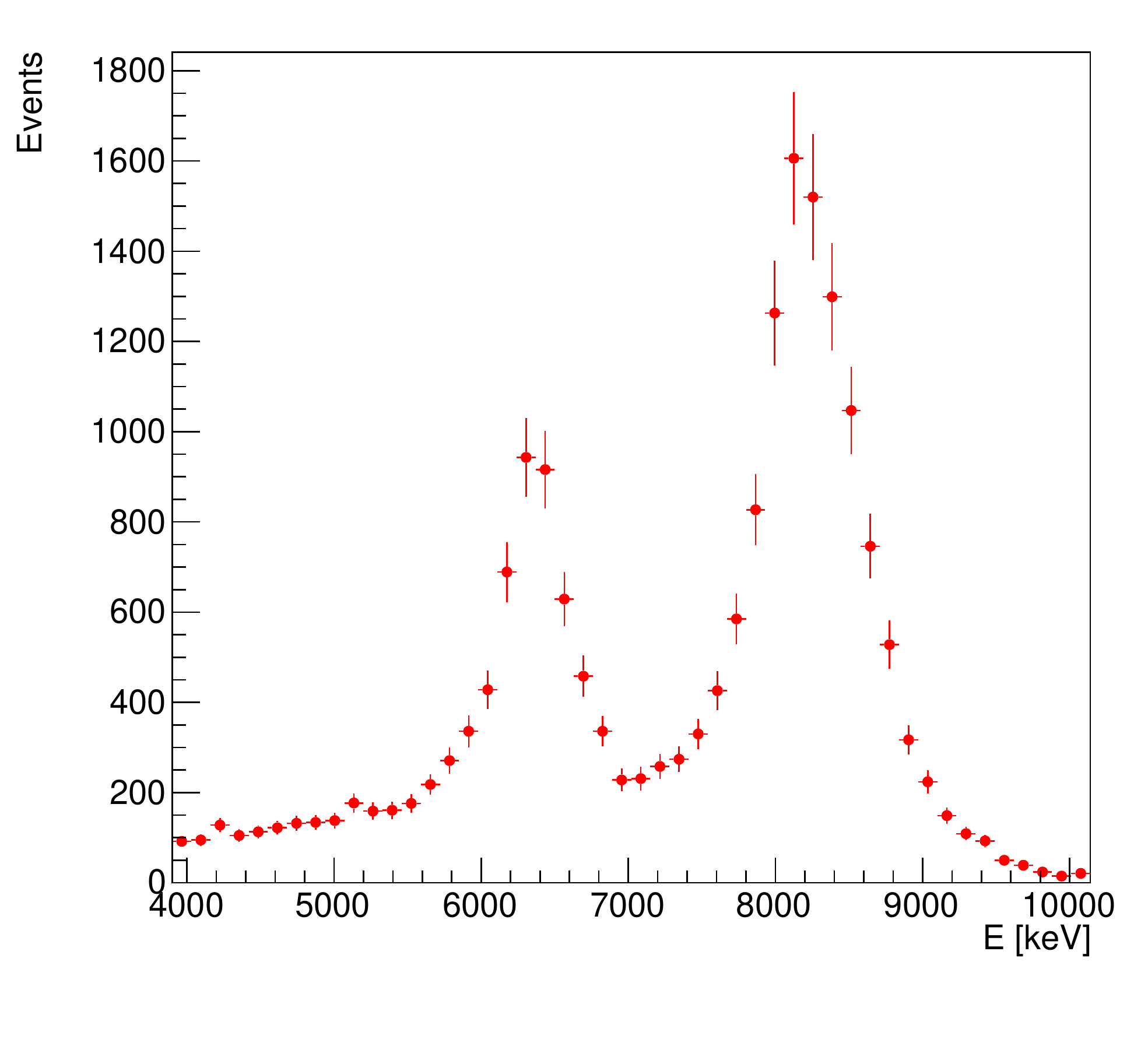}  }
\caption{Energy spectrum of particles as measured by the MX-10 camera, recorded in a natural cave. Two strong alpha peaks at energies of approximately 6.4 and 8.2 MeV are clearly seen, while their exact position is a subject to imperfect calibration of the particular device used.}  
\label{fig:alpha:spect}
\end{figure}

Some basic facts have been clarified, known perhaps to most of cave climatologists or radon specialists,
but it was found that the energy peaks of the observed alpha particle are consistent with those from polonium isotopes instead of the direct radon origin, i.e. the observed alpha particles seen come from solid decay products of radon, probably adhered in a form of aerosols directly on the chip. This is confirmed by the relatively sharp peaks at energies of about 6.4 and 8.2~MeV of alpha spectral particles as displayed in Fig.~\ref{fig:alpha:spect}, corresponding probably to alpha particles from Po-218 and Po-214 decays, respectively, with an indication of the camera overestimating the energy by about 7\% (the actual energies are about 6.0 and 7.7~MeV~\cite{periodic}).


\clearpage
\section{Particle camera aboard a commercial airplane}

\begin{figure}[!h]
\centerline{ 
  \includegraphics[width=0.450\textwidth,height=0.450\textwidth]{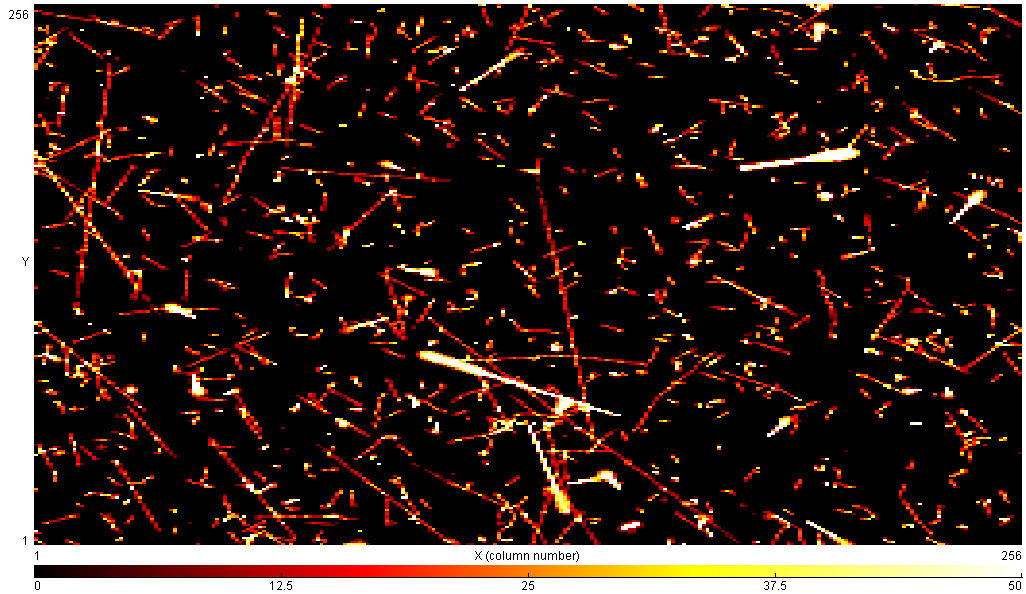}
  \includegraphics[width=0.450\textwidth,height=0.450\textwidth]{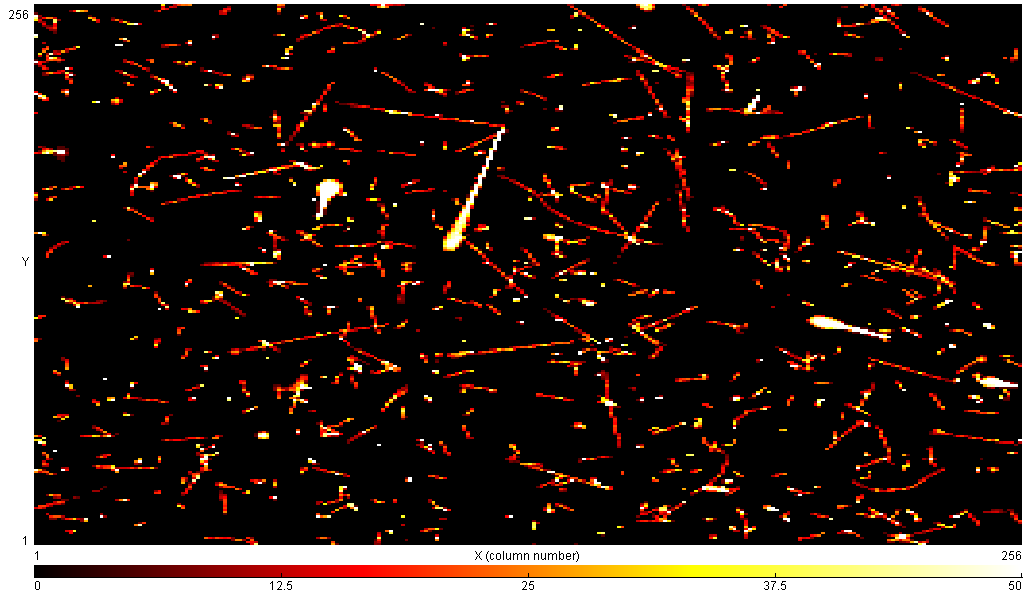}
}
  \caption{Example of 15 min exposures recorded aboard a commercial air plane over Europe at altitude of about 10~km.}
\label{fig:plane}
\end{figure}

\begin{figure}[!h]
  \centerline{
    \includegraphics[width=0.750\textwidth]{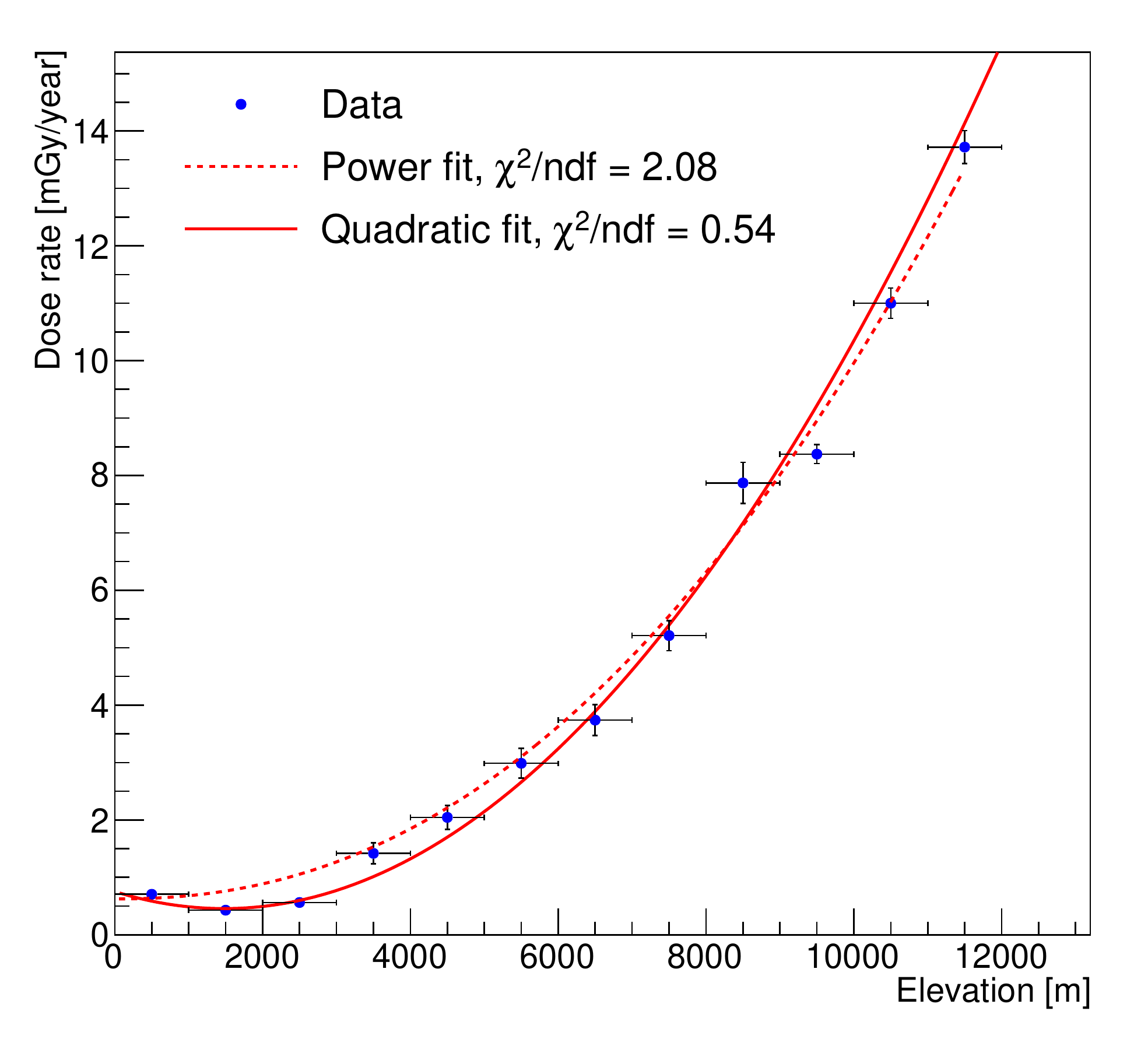}  }
  \caption{The dose rate recorded by the particle camera silicon chip in mGy/year as function of the altitude above the sea level in meters. The solid (dashed) line is a quadratic (power) fit to data. Error bars are statistical only, the standard $\chi^2$ divided by the number of degrees of freedom (ndf) is indicated in plot legend.}
\label{fig:dose}
\end{figure}

As the camera works as a simple USB device, measurements with a laptop can be carried aboard a commercial airplane.
Fig.~\ref{fig:plane} which shows examples of 15 min exposures at altitude of about 10 km.
In addition, pairing this information with data from an external GPS (in a digital camera) with a synchronized clock, the radiation level was studied as a function of the altitude~\cite{SOC}.
In Fig.~\ref{fig:dose} one can observe the effect of initially reduced radiation with altitude at about 1~km, followed by increase from cosmic rays, as observed originally also by V.~Hess and most notably by W.~Kolh\"{o}rster~\cite{cosmic_wiki,FICK201450}. Error bars are statistical only, accounting for the fluctuation in the number of observed particles in frames included to the sum of the observed energy in given altitude range.
Thanks to the pixelized detector, one can clearly see many straight tracks of muons, as well as thick tracks of highly ionizing particles of energies up to 20 MeV, i.e. surpassing natural sources of radiation (alpha particles from natural radioisotopes have energies up to about 8~MeV).
All these are in sharp contrast to radiation levels and patters observed at the Earth surface. The dose recorded by the camera is about 15 times higher at altitude of 10~km, compared to the altitude of the laboratory of about 200~m.


\begin{figure}[!h]
\centerline{ 
  \includegraphics[width=0.450\textwidth,height=0.450\textwidth]{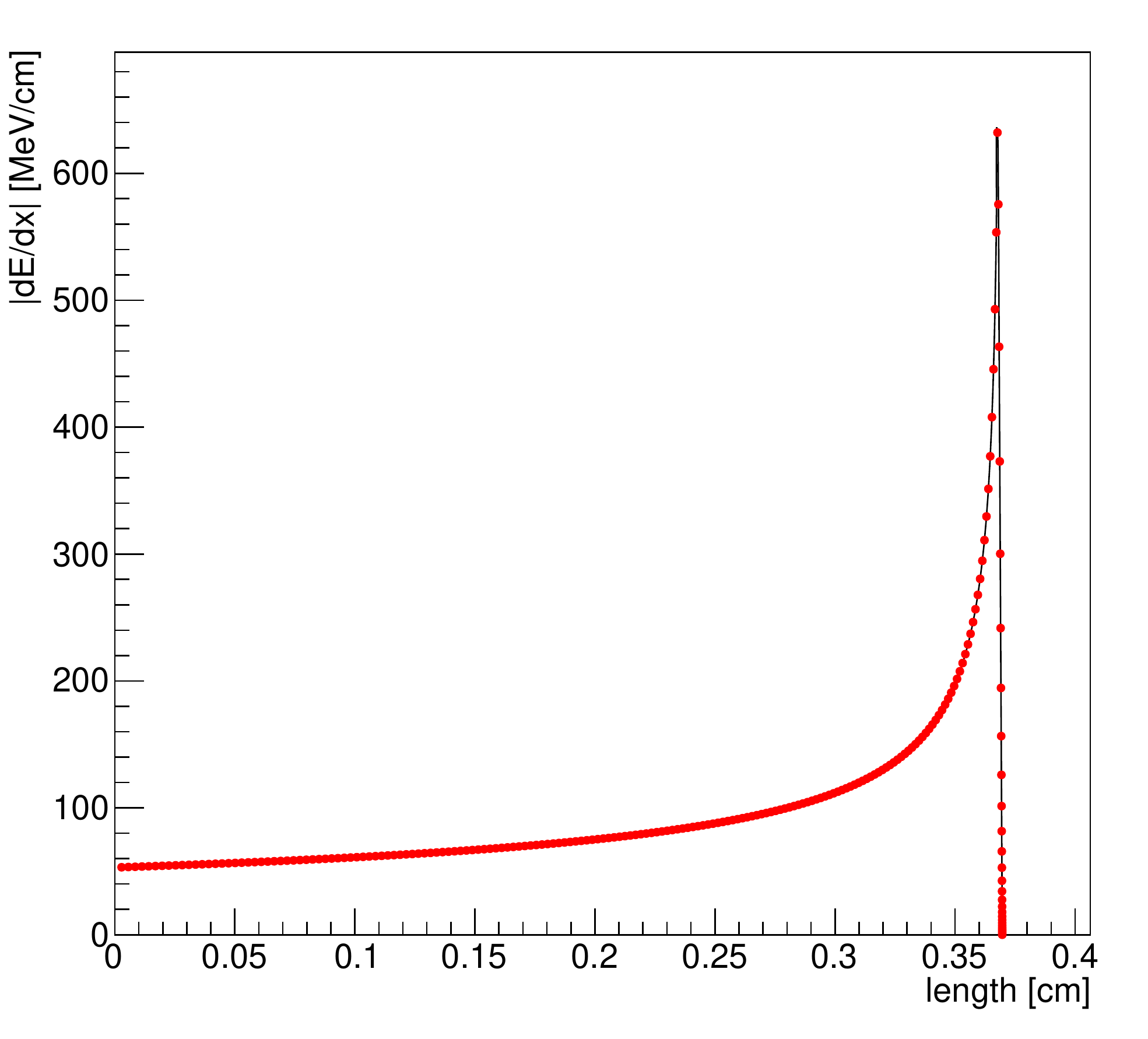}
  \includegraphics[width=0.450\textwidth,height=0.450\textwidth]{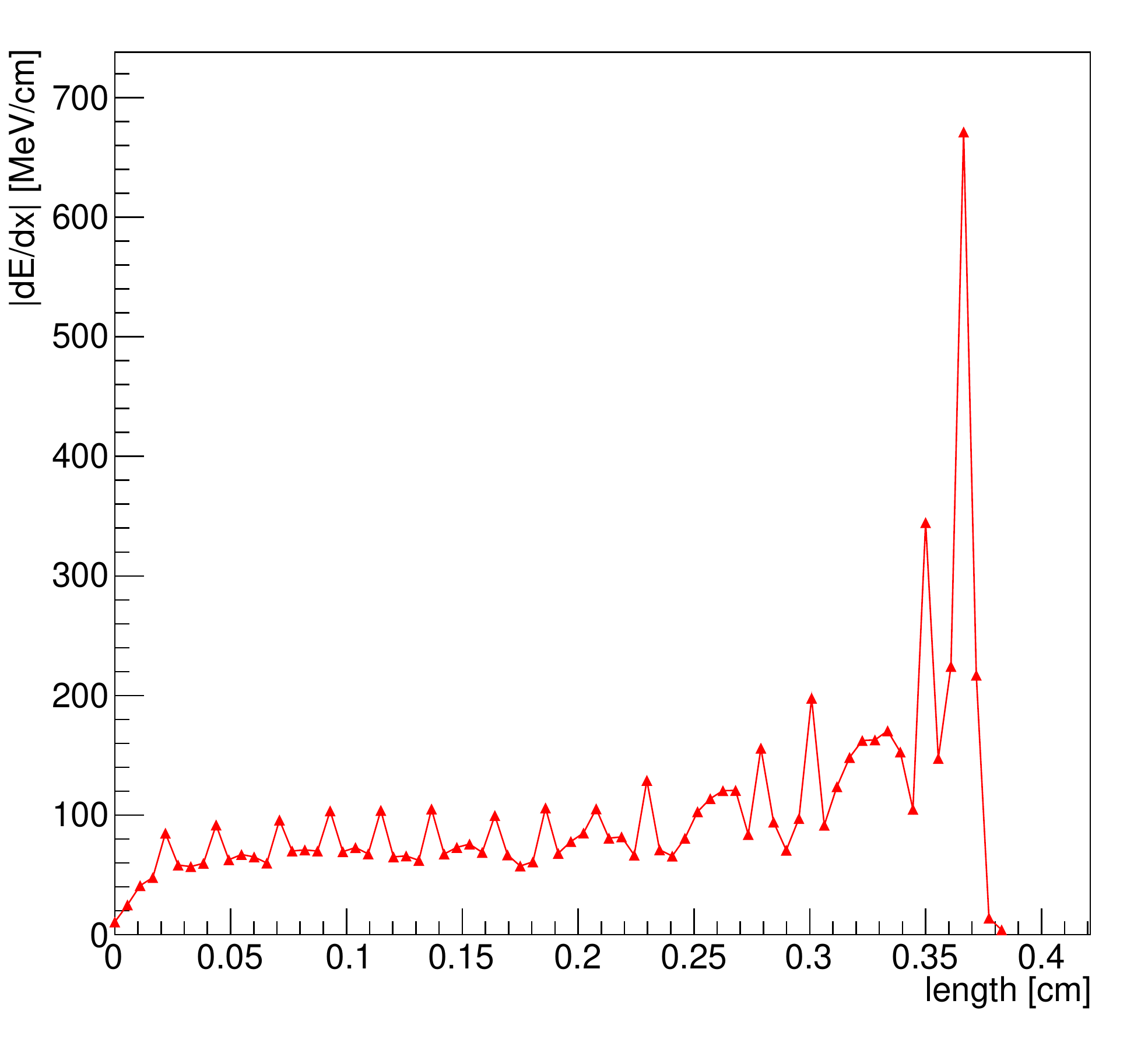}
}
\caption{Simulated Bragg peak (left), i.e. the enhancement of the energy loss per track length towards the end of a trajectory, of a 33~MeV deuteron in silicon, 
  and an example of a recorded 40 MeV track from flight data in MX-10 camera (right) exhibiting a similar pattern.}
\label{fig:bragg}
\end{figure}

Tracks of energies above 10~MeV were observed, with extreme cases of energies of~21 and even 40~MeV, the most energetic one carrying a sign of a Bragg peak, i.e. sharp increase of energy loss towards the end of the trajectory, followed by a rapid decrease, see Fig.~\ref{fig:bragg}, which has an important application in medicine in hadron therapy. The observed range and losses pattern is similar to predicted losses of a deuteron in silicon with initial kinetic energy lower by 20\% compared to the observed track. The reasonable agreement of the measured and predicted curves was reached by the change in the energy, motivated by the camera overcalibration, and also by the adopted model of losses below the application limit of the Bethe-Bloch formula~\cite{pdg2} which was chosen to correspond to the particle velocity of $0.03c$, below which the energy losses were linearly interpolated to zero at rest.

\begin{figure}[!h]
\centerline{ 
  \includegraphics[width=0.450\textwidth,height=0.450\textwidth]{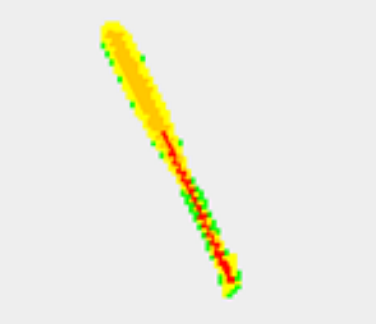}
  \includegraphics[width=0.450\textwidth,height=0.450\textwidth]{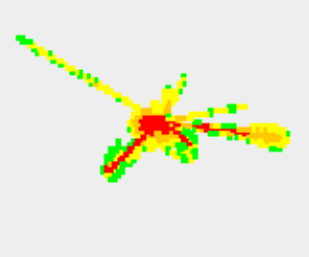}
}
\caption{The 40 MeV track candidate (left) exhibiting the Bragg peak and a nuclear interaction of total of 77 MeV deposited (right). The color coding of the pixels is as follows: 0--10 keV: green, 10--50 keV: yellow, 50--150 keV: orange, $\geq 150$ keV: red.}
\label{fig:candidates}
\end{figure}

In addition, one nuclear interaction was identified, with total of 77 MeV deposited into the device (see Fig.~\ref{fig:candidates}).


\section{Particle camera at an accelerator}

\begin{figure}[!h]
 \centerline{ 
 \includegraphics[width=0.450\textwidth,height=0.450\textwidth]{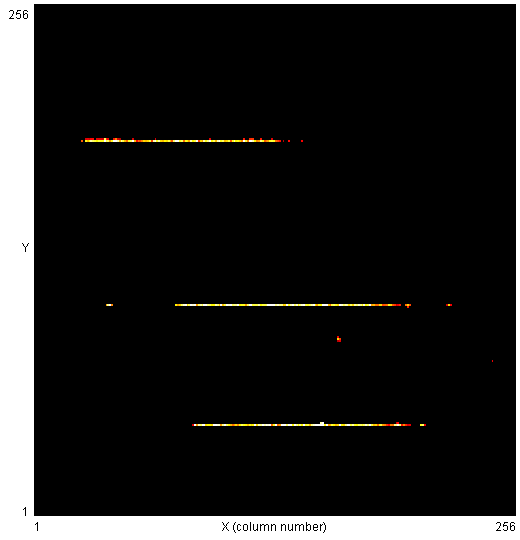}
 \includegraphics[width=0.450\textwidth,height=0.450\textwidth]{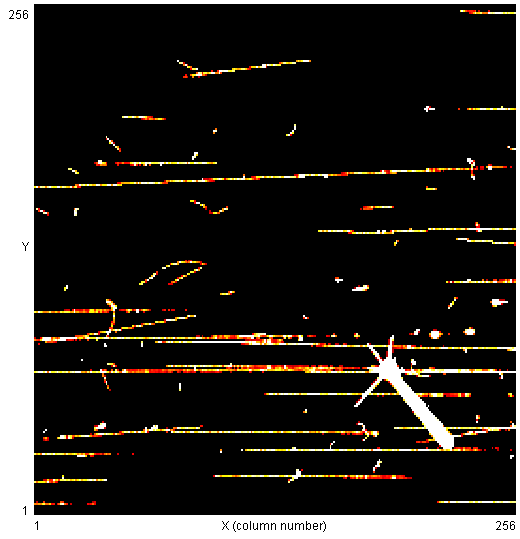}
}
  \caption{Example of frames recorded by the particle camera with the chip inserted parallel to a beam of muons (left) and charged pions (right) at the SPS test beam area at CERN.}
\label{fig:TB}
\end{figure}

In Fig.~\ref{fig:TB} one can see example frames recorded by the particle camera with the chip inserted parallel to a beam of muons (left) and charged pions (right) at the SPS test beam area at CERN. The figure clearly demonstrates the complexity of hadronic interactions of pions compared to muons.
In addition, it is not uncommon to see actually a break-up of a nucleus into several heavy-ionizing fragments, i.e. the alchemists' dream of changing a chemical element into another one.

\section{Discussion}

Students have performed and contributed in a major way to interesting experiments with various sources of the ionization radiation using the particle camera MX-10. Ranging from simple background measurements to the muon direction analysis, coming to radon survey in caves, the study of cosmic rays, dose variation with altitude, analysis of tracks of exotic heavily-ionizing particles, and finally to an example of the detector as a detection element in a fixed-target setup at an accelerator.

\subsection{Responsibility}

Already this high-school level research brings many questions on how to present and explain results to general public.
The issue is e.g. how to present (the interesting yet harmless for a mere visitor) elevated radiation background in the caves environment, in order not to disseminate confusion or even fear of ionization radiation levels on a plain or in caves, while bringing these exciting facts to interested students, readers, enthusiasts.
It is a sad truth that anything related to radon or ionizing radiation sparks often negative emotions in general public, although we are all being exposed to small harmless doses of radiation from many natural sources, including radon, medical imaging, food and airplane travel; and depending on the geographic location on Earth, not only as a function of the altitude above the see level but also as function of the Earth's crust local composition.

Measurements were first discussed in a local high-school journal in a form of an interview and then also in a local TV station, later, without letting authors and the supervisor properly know, appearing also in internet news, catching attention of the management of the caves, who were not particularly happy seeing the discussion of radon-related radiation in the caves (which is of course carefully being monitored in over long term periods). Thus, through a primarily science project, students actually found themselves be taught a lesson on the interaction with media and on the way how to communicate their results.

Nevertheless, we are responsible to the authorities and the public for their support and we have the social responsibility to spread knowledge and education. It is the duty of us, scientists, to present and explain facts in a way which is understandable to all. This can be particularly well achieved by attracting students to related projects already at the high-school level who can then present their findings to their schoolmates and who can then share knowledge in the most effective way, i.e. by themselves, directly within their age group. Last, we prove that high-school students can contribute to the process of writing a journal article containing their own work.


\section{Acknowledgments}

This educational research was performed using the MX-10 device by the company JABLOTRON ALARMS, equipped by the Medipix/Timepix chip~\cite{VYKYDAL2006112} and software Pixelman~\cite{HOLY2006254} by IEAP, Czech Technical University, Prague, Czech Republic.
Data were analyzed by a private code in C++, Python and using libraries of the ROOT analysis framework~\cite{Brun:1997pa}; and using Java, MS Excel and Visual Basic macros for the case of processing the flight data.
We thank T.~S\'{y}kora for bringing our attention to the MX-10 camera and to the Hough transformation, and to L.~Chytka for recording the data with MX-10 in his spare time during the SPS test beam campaign for the time-of-flight detector for the ATLAS Forward Proton detector.
We also thank P. Baro\v{n} for the interpretation of the alpha spectra.
Our thanks belong to the Nature Conservation Agency of the Czech Republic and the staff of the Zbra\v{s}ov Aragonite Caves, Teplice nad Be\v{c}vou, Czech Republic, for providing us with the opportunity to perform measurements over several nights in the cave system.
Last, the students would like to thank the pedagogical staff of the Grammar school in B\'{\i}lovec; and M.~Komsa, L.~Balc\'{a}rek from the Grammar school in Uni\v{c}ov and R.~D\v{e}rda from the Technical and training school in Uni\v{c}ov for their support and for allowing them to travel to Palack\'{y} University in Olomouc in order to pursue their high school science projects. 
D.S. and J.P. took the third place in the physics category of the national competition of Czech high school science projects (SO\v{C}) in 2018.

J.K. gratefully acknowledges the support by the Operational Programme Research, Development and Education -- European Regional Development Fund, project \\ no.~CZ.02.1.01/0.0/0.0/16\_019/0000754 of the Ministry of Education, Youth and Sports (MSMT) of the Czech Republic; and support from the grant LTT17018 of MSMT, Czech Republic.

\bibliography{main}{}
\bibliographystyle{plain}

\end{document}